\documentclass[10pt]{ismd08}
\usepackage{graphicx}
\usepackage{cite,./mcite}

\begin{document}
\title{Exclusive photoproduction of dileptons at high energies}
\author{M.V.T. Machado\protect\footnote{\ \ speaker}}
\institute{Centro de Ci\^encias Exatas e Tecnol\'ogicas, Universidade Federal do Pampa \\
Campus de Bag\'e, Rua Carlos Barbosa. CEP 96400-970. Bag\'e, RS, Brazil}
\maketitle
\begin{abstract}
The exclusive photoproduction of lepton pairs on nucleon and nucleus target is investigated within the  high energy color dipole approach, where the main physical quantity is the dipole-target elastic scattering amplitude that captures the main features of the dependence on atomic number $A$, on energy and on momentum transfer $t$. These calculations are input in predictions for electromagnetic interactions in $pp$ and $AA$ collisions to be measured at the LHC.

\end{abstract}

The physics of large impact parameter interactions \cite{YRUPC} at the LHC and Tevatron has raised great interest as the these electromagnetic interactions in $pp$ and $AA$ collisions extend the physics program of photon induced processes beyond the energies currently reached at DESY-HERA. This can happen in a
purely electromagnetic process through a two-photon interactions or in an
interaction between a photon from one of the nuclei and the other ``target'' nucleus. These ultraperipheral collisions (UPCs) are a good place to constraint the photonuclear cross sections as the dominant processes in UPCs are photon-nucleon (nucleus) interactions. Electromagnetic interactions can also be studied with beams of
protons or anti-protons, but there is then no $Z^2$-enhancement in the
photon flux in contrast to $AA$ collisions. Several analysis are currently being done at Tevatron focusing on such processes. For instance, CDF Collaboration is analyzing the exclusive production of muon pairs,
$p \overline{p} \rightarrow p \overline{p} + \mu^+ \mu^-$, at lower invariant
masses \cite{Confpress}. The two main contributions to these events are, as with
heavy-ion beams, $\gamma \gamma \rightarrow \mu^+ + \mu^-$ and
$\gamma + I\!P\rightarrow J/\Psi (\mathrm{or} \,\Psi') $, followed by the meson decay
into a dilepton pair.

In this contribution, we summarize the results presented in Ref. \cite{Magdilep}, where the high energy color dipole approach \cite{dipole} is used to study the exclusive photoproduction of lepton pairs. The motivation to consider the color dipole formalism is due to the fact that the electromagnetic deeply virtual Compton scattering (DVCS) cross section at high energies is nicely reproduced in several implementations of the dipole cross section at low $x$ \cite{KMW,MPS,Watt,MW}. The present process at small $t$ and large timelike virtuality of the outgoing photon shares many features of DVCS.  Simple models for the elementary dipole-hadron scattering amplitude that captures main features of the dependence on atomic number $A$, on energy and on momentum transfer $t$ were considered. Our investigation is complementary to conventional partonic description of timelike Compton scattering (TCS) \cite{Diehl}, $\gamma p \rightarrow \gamma^*+p$, which considers the relevant generalized  parton distributions (GPDs).  It should be noticed that the TCS process has so far only been studied at LO in the collinear factorization framework in terms of the quark GPDs and sub-processes initiated by gluons have not been considered \cite{Diehl}. This approach has been recently considered to make prediction for the relevant kinematics for LHC \cite{Pire1,Pire2} and detailed investigation of competing processes (like Bethe-Heitler contribution) and possible interference term is presented.

\section{TCS process within the color dipole approach}

In the color dipole picture \cite{dipole}, the scattering process $\gamma p\rightarrow \gamma^*p$ is assumed to proceed in three stages: first the incoming real photon fluctuates into a quark--antiquark pair, then the $q\bar{q}$ pair scatters elastically on the proton, and finally the $q\bar{q}$ pair recombines to form a virtual photon (which subsequently decays into lepton pairs). The amplitude for production of the exclusive final state such as a virtual photon in TCS, is given by \cite{MPS,KMW,MW}
\begin{eqnarray}
 \mathcal{A}^{\gamma p\rightarrow \gamma^* p}(x,Q,\Delta)  =  \sum_f \sum_{h,\bar h} \int\!d^2\vec{r}\,\int_0^1\!d{z}\,\Psi^*_{h\bar h}(r,z,Q)\,\mathcal{A}_{q\bar q}(x,r,\Delta)\,\Psi_{h\bar h}(r,z,0)\,,
  \label{eq:ampvecm}
\end{eqnarray}
where $\Psi_{h\bar h}(r,z,Q)$ denotes the amplitude for a photon to fluctuate into a quark--antiquark dipole with helicities $h$ and $\bar h$ and flavour $f$. The quantity $\mathcal{A}_{q\bar q}(x,r,\Delta)$ is the elementary amplitude for the scattering of a dipole of size $\vec{r}$ on the proton, $\vec{\Delta}$ denotes the transverse momentum lost by the outgoing proton (with $t=-\Delta^2$), $x$ is the Bjorken variable and $Q^2$ is the photon virtuality.

As one has a real photon at the initial state, only the transversely polarized overlap function contributes to the cross section.  The expression for the  overlap function can be found for instance in Ref. \cite{MW}. In our numerical calculations we use still the space-like kinematics and we expect not a large deviation from the correct kinematics. However, the approximation we have considered to estimate the cross section should be taken with due care.  It should be noticed that some corrections to this exclusive process are needed. For TCS one should use the off-diagonal gluon distribution, since the exchanged gluons carry different fractions $x$ and $x^\prime$ of the proton's (light-cone) momentum. The skewed effect can be accounted for, in the limit that $x^\prime \ll x \ll 1$, by multiplying the elastic differential cross section by a  correction factor \cite{Shuvaev:1999ce}. We quote Ref. \cite{Magdilep} for details.

In the numerical calculations we consider the non-forward saturation model of Ref. \cite{MPS} (MPS model), which has the advantage of giving directly the $t$ dependence of elastic differential cross section without the necessity of considerations about the impact parameter details of the process. It is based on the studies about the growth of the dipole amplitude towards the saturation regime is the geometric scaling regime \cite{gsincl}.  The geometric scaling property can be extended to the case
of non zero momentum transfer \cite{MAPESO}, provided $r\Delta\ll 1$, where the elementary dipole amplitude now reads as:
\begin{eqnarray}
\label{sigdipt}
\mathcal{A}_{q\bar q}(x,r,\Delta)= 2\pi R_p^2\,e^{-B|t|}N \left(rQ_{\mathrm{sat}}(x,|t|),x\right),
\end{eqnarray}
with the $t$ dependence of the saturation scale being parametrised as
\begin{eqnarray}
\label{qsatt}
Q_{\mathrm{sat}}^2\,(x,|t|)=Q_0^2(1+c|t|)\:\left(\frac{1}{x}\right)^{\lambda}\,, \end{eqnarray}
in order to interpolate smoothly between the small and intermediate transfer
regions. The scaling function $N$ is obtained from the forward saturation model
\cite{Iancu:2003ge}.

In the case of nuclear targets, we make use of studies of Ref. \cite{Armesto_scal} where the high energy $l^{\pm}p$,  $pA$ and $AA$ collisions have been related through geometric scaling. In that approach, the nuclear saturation scale was assumed to rise with the quotient of the transverse parton densities, $\kappa_A = A\pi R_p^2/(\pi R_A^2)$,  to the power $\Delta \approx 1$, that is  $Q_{\mathrm{sat},A}^2=\kappa_A^{\Delta}\,Q_{\mathrm{sat},p}^2$ ($R_A$ is the nuclear radius). This assumption successfully describes small-$x$ data for $ep$ and $eA$ scattering using $\Delta =1.26$ and a same scaling curve for the proton and nucleus \cite{Armesto_scal}. Therefore, we propose the following expression for the elementary dipole amplitude for nuclear case:
\begin{eqnarray}
\label{sigdipta}
\mathcal{A}_{q\bar q}(x,r,\Delta;A)= 2\pi R_A^2\,F_A(t)\,N \left(rQ_{\mathrm{sat},A}(x,|t|),x\right),
\end{eqnarray}
where $F_A(t)$ is the nuclear form factor. It should be stressed that we consider only the coherent nuclear scattering, whereas the incoherent case is neglected.

As we investigate the exclusive photoproduction of a heavy timelike photon which decays into a lepton pair, $\gamma p \rightarrow \ell^+\ell^-p$. Therefore, for the $\ell^+\ell^-$ invariant mass distribution from the virtual $\gamma^*$ decay we have (with $Q^2=M_{\ell^+\ell^-}^2$),
\begin{eqnarray}
 \frac{d\sigma }{dM_{\ell^+\ell^-}^2}\left(\gamma p\rightarrow \ell^+\ell^- p\right) = \frac{\alpha_{em}}{3\pi M_{\ell^+\ell^-}^2}\,\sigma\left(\gamma p \rightarrow \gamma^* p \right).
\end{eqnarray}

\section{Results and Conclusions}

\begin{figure}[t]
\centerline{\includegraphics[scale=0.43]{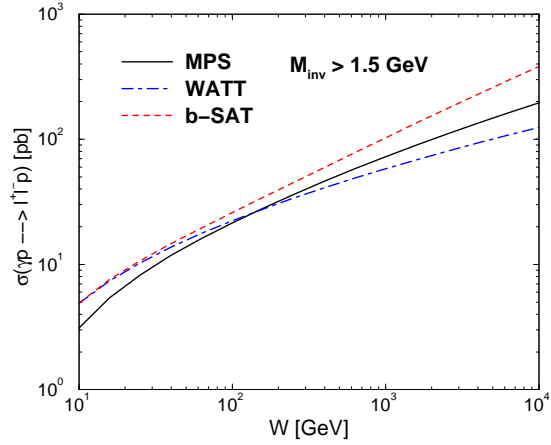}}
\caption{The integrated cross section ($M_{\ell^+\ell^-}\geq 1.5$ GeV and $|t|\leq 1$ GeV$^2$) as a function of photon-nucleus centre of mass energy.}
\label{fig:1}
\end{figure}

In Fig. \ref{fig:1} we show the total cross section, $\sigma(\gamma p \rightarrow \ell^+\ell^-\,p)$, integrated over dilepton invariant mass $M_{\ell^+\ell^-}\geq 1.5 $ GeV. In this plot, we compare the MPS model (solid line) with other implementations of the elementary dipole-nucleon scattering amplitude. For energies $W>>100$ GeV, a power fit can be performed in the form $\sigma = A\,(W/W_0)^\alpha$, with $W_0=1$ GeV. Considering the MPS model, one obtains the values $A= 3$ pb and $\alpha = 0.46$. In order to study the sensitivity to the model dependence, we compare the MPS saturation model to two different distinct saturation models. The first one is the recent implementation of the impact parameter Color Glass Condensate model \cite{Watt} (hereafter WATT, dot-dashed line). The second one is the impact parameter saturation model \cite{KMW}(hereafter b-SAT, long dashed line). In these implementations, the elementary dipole-nucleon scattering amplitude is written in the impact parameter space. At high energies, a power fit can be also performed for the WATT and b-SAT models. For WATT impact parameter model, we obtain the parameters $A= 4.3$ pb and $\alpha = 0.37$. In the b-SAT model we have the values $A= 2$ pb and $\alpha = 0.57$. We clearly verify a distinct energy dependence, which depends on the characteristics features of the phenomenological models.  The main point is the value of the parameter $\lambda$ entering at the saturation scale. The WATT model gives the softer energy dependence, which comes from the small value of $\lambda = 0.12$ in the saturation scale, $Q_{\mathrm{sat}}^2(x,t=0)\propto x^{\lambda}$. On  the other hand, in the MPS model one has $\lambda = 0.22$ (a factor two larger than in WATT model) and the QCD evolution makes the effective $\lambda$ value for b-SAT model to be large. The different overall normalization at energies $W\leq 100$ GeV for the MPS model is probably result of distinct behavior towards low energies.

Concerning the nuclear targets, in Ref. \cite{Magdilep} we also compute the integrated cross section per nucleon as a function of energy for $A=208$ (Lead), which is relevant for electromagnetic interactions at $AA$ collisions at the LHC. The nuclear version of MPS model at high energies ($W\geq 100$ GeV) can be parameterized as $\sigma_{\mathrm{MPS}}(\gamma A\rightarrow \ell^+\ell^-\,A)=6.1\mathrm{pb}\,(W/W_0)^{0.39}$. For the b-SAT model one obtains at high energies   $\sigma_{\mathrm{bSAT}}(\gamma A\rightarrow \ell^+\ell^-\,A)=5\mathrm{pb}\,(W/W_0)^{0.51}$. Accordingly, we verify a larger suppression in the MPS model than in b-SAT, which is directly related to the nuclear saturation scale at each model. Absorption is evident in MPS model, where the effective power on energy has diminished in the nuclear case.

The present calculations are input for the exclusive photoproduction of dileptons in electromagnetic interactions in $pp$ and nucleus-nucleus collisions. Let us concentrate on the $pp$ case, where the processes is characterized by the photon - proton interaction, with the photon stemming from the electromagnetic field
of one of the two colliding hadrons.  The total cross section for the $p\,p\rightarrow p\otimes \ell^+\ell^- \otimes p$ process is obtained by the product of the photon-proton cross section and the photon energy spectrum, $dN/d\omega$, and integration over the photon energy, $\omega $:
\begin{eqnarray}
\sigma\,(pp\rightarrow p+\ell^+\ell^-+p)=2\int_0^{\infty}d\omega\frac{dN_{\gamma}}{d\omega}\,\sigma\,(\gamma p\rightarrow \ell^+\ell^-+p)\,
\label{sigAA}
\end{eqnarray}
where $\gamma_L = \sqrt{s_{pp}}/2m_p$ is the Lorentz boost  of a single beam,  $W_{\gamma p}^2\approx 2\omega\sqrt{s_{pp}}$  and
$\sqrt{s_{pp}}$ is  the centre of mass energy of the
hadron-hadron system \cite{YRUPC}. The initial factor of 2 in above equation accounts for the interchange of the photon emitter and the target. This process is characterized by small momentum transfer and energy loss, which implies that the outgoing hadrons should be detected in the forward regions of detectors. The final state is relatively clear, presenting two rapidity gaps and central dilepton production. Therefore, this process has a similar final state as the produced in gamma-gamma scattering. In Table 1, we present  an estimative for $pp$ collisions for Tevatron and the LHC energies (integrated over invariant mass $M_{\mathrm{inv}}>1.2$ GeV).  For the Tevatron case, we also present the result for the current cut on invariant dilepton mass $3 \,\mathrm{GeV} \leq M_{\ell^+\ell^-} \leq 4 \,\mathrm{GeV}$ which refers to upcoming exclusive  dimuon CDF measurements \cite{Albrow}. In order to compare it to the QED case, we quote the value $\sigma (|\eta|<0.6,3<M_{\mu^+\mu^-}<4\,GeV)=2.18$ pb from the process $\gamma \gamma \rightarrow \mu^+\mu^-$.  In the $AA$ the photon flux is enhanced by a factor $\propto Z^2$, then we would expect the cross sections to reach dozens of nanobarns for $PbPb$ collisions at the LHC. The result presented here can be compared to QCD factorization formalism for exclusive processes involving the GPDs \cite{Pire2}. The authors in Ref \cite{Pire2} have found a strong dependence on the factorization scale. The Compton contribution was estimated to be 1.9 pb at LHC using the cut $2.12 \leq M_{\ell^+\ell^-}\leq 2.35$ GeV and additional cuts on polar and azimuthal angles. Using the same kinematical cut they found a cross section of 2.9 pb for the Bethe-Heitler contribution. We believe that applying similar cuts, our results for the cross section can be compatible with values presented in Ref. \cite{Pire2} for the LHC case.

As a summary, using the color dipole formalism we studied the timelike Compton scattering. Such an approach is robust in describing a wide class of exclusive processes measured at DESY-HERA and at the experiment CLAS (Jeferson Lab.), like meson production, diffractive DIS and DVCS. Our investigation is complementary to conventional partonic description of TCS, which considers quark handbag diagrams (leading order in $\alpha_s$) and simple models of the relevant GPDs. In particular, the results could be compared to pQCD diagrams involving gluon distributions which are currently unknown. Using current phenomenology for the elementary dipole-hadron scattering, we estimate the order of magnitude of the exclusive photoproduction of lepton pairs.   These calculations are input in electromagnetic interactions in $pp$ and $AA$ collisions to measured at the LHC. We found that the exclusive photoproduction of lepton pairs in such reactions should be sizable.

\begin{table}[t]
\begin{center}
\begin{tabular}{|l|c|c|}
\hline
$\sqrt{s_{pp}}$   & $M_{\ell^+\ell^-}> 1.2$ GeV  &  3 GeV $\leq M_{\ell^+\ell^-} \leq$ 4 GeV  \\
\hline
1.96 TeV &  7 pb  & 0.4 pb\\
\hline
14  TeV &  25 pb  &  --- \\
\hline
\end{tabular}
\end{center}
\caption{Cross section for exclusive dilepton photoproduction at Tevatron and LHC.}
\label{tab:table1}
\end{table}

\begin{footnotesize}
\bibliographystyle{ismd08}

\end{footnotesize}
\end{document}